# Electric field dependent thermal conductivity of relaxor ferroelectric PMN-33PT through changes in the phonon spectrum

Delaram Rashadfar,[a] Brandi L. Wooten [b,*] and Joseph P. Heremans [a,b,c,*]

In ferroelectric materials, an electric field has been shown to change the phonon dispersion sufficiently to alter the lattice thermal conductivity, opening the possibility that a heat gradient could drive a polarization flux, and technologically, also opening a pathway towards voltage-driven, all solid-state heat switching. In this report, we confirm the validity of the theory originally developed for $Pb(Zr,Ti)O_3$ (PZT) on the ferroelectric relaxor $0.67Pb[Mg_{1/3}Nb_{2/3}]O_3$-$0.33PbTiO_3$ (PMN-33PT). In the theory, the change in sound velocity and thermal conductivity with electric field relates to the piezoelectric coefficients and the Grüneisen parameter. It predicts that in PMN-33PT the effect should be an order of magnitude larger, and of opposite sign as in PZT; this is confirmed here experimentally. The effects are measured on samples never poled before and on samples that underwent multiple field sweep cycles and passed through two phase transitions with change in temperature. The thermal conductivity changes are linked to variations in the piezoelectric coefficients and can be as large as 8-11% at $T \geq 300$ K. To date, this is the only means of heat conduction modulation that utilizes changes in the phonon spectrum. While this technology is in its infancy, it offers another path to future active thermal conduction control.

## Introduction

Long-range order occurs in materials that are ferromagnetic (FM) or ferroelectric (FE). The interactions between atomic magnetic or electric dipoles lead to a collective magnetization or polarization moment ($M/P$) that exhibits a hysteresis loop in response to a magnetic or electric field ($H/E$). At higher fields, dipole moments align with the field direction, achieving saturation at moments ($M_S/P_S$). At intermediate field strengths, the materials create domains where $M/P$ are oriented in various directions. As the temperature rises, $M_S/P_S$ diminish due to increased thermal fluctuations. Additionally, there exists a non-zero pyroelectric coefficient $\Pi_{PY} \equiv {}^{d|P|}/_{dT}$ at all field strengths. Both factors indicate that a temperature gradient ($\nabla T$) produces a polarization gradient. In irreversible thermodynamics, the changes in extensive quantities such as heat ($Q$) and magnetization or polarization are linked to a heat flux $j_Q$ accompanied by a spin flux $j_S$ in ferromagnets, and its extension to ferroelectrics the polarization flux $j_P$, allowing for the expression of polarization Onsager relations [1]:

$$-j_P = \sigma \nabla E - S\sigma \nabla T \quad (1)$$

$$j_Q = \sigma \Pi \nabla E - \kappa \nabla T \quad (2)$$

where σ is the polarization conductivity tensor, and $S$, $\Pi$, and $\kappa$ represent the polarization Seebeck, Peltier, and thermal conductivity tensors, respectively.

For FMs, the thermal fluctuations of magnetization involve spin waves or magnons, while in FEs, it relates to a specific type of phonon called ferrons, as identified by Bauer et al. [1] Like how $M_S$ decreases with rising temperature due to an increase in magnon population, $P_S$ follows suit due to a higher density of ferrons. Intuitively, in displacement ferroelectrics, ferrons include optical phonons that involve the displacement of the ions that give rise to the atomic dipoles. However, the distinction between acoustic and optical modes remains well-defined only at zero wavevector ($k$); at non-zero $k$, the optical and acoustic phonons couple due to strain. Ferrons are thus optical and acoustic modes strain-coupled by the piezoelectric coefficients. In principle, one then expects an electric field dependence of the phonon dispersion and hence the lattice thermal conductivity. The former effect is observed in $0.70Pb[Mg_{1/3}Nb_{2/3}]O_3$-$0.30PbTiO_3$ (PMN-30PT) [2] and the latter in $Pb(Zr,Ti)O_3$ (PZT) [3] where the field is shown to affect the lattice thermal conductivity by about 2%. The present article strongly amplifies the earlier work of Wooten et al. [3] by showing effects one and two orders of magnitude larger in the field dependent thermal conductivity of PMN-33PT.

A quantitative theory [3] for the dependence of sound velocity, lattice thermal conductivity and diffusivity with no adjustable parameters relates changes in the sound velocity $v$ to the piezoelectric coefficients $d_{33}$ and $d_{31}$ and the Grüneisen parameter $\gamma$. The piezoelectric coefficients are defined as $d_{33} = {}^{\partial e_{33}}/_{\partial E}$ and $d_{31} = {}^{\partial e_{11}}/_{\partial E}$, where $e_{33(11)}$ is the strain tensor component that gives the compression or expansion along (perpendicular to) the direction of ferroelectric order, which is parallel to the applied field $E = E_3$. The mode ($i$) and wavevector ($k$) dependent Grüneisen parameter is the logarithmic derivative of that phonon mode's frequency vis-à-vis the volume $V$ of the sample, $\gamma(i,k) \equiv {dln(\omega(i,k))}/{dln(V)}$, by definition. For longitudinal or transverse acoustic phonons at low $k$, ($i$= LA or TA, respectively), this gives $\gamma(i) \equiv {dln(v(i))}/{dln(V)}$. The basis of the theory is to first describe how the piezoelectric coefficients relate the change in volume of the sample with applied field to $(d_{33}+2d_{31})$ and then relate that to the logarithmic derivative of the sound velocity for each mode:

$$\frac{v'}{v_0} = \frac{dln(v)}{dE} = -\gamma(d_{33}+2d_{31}). \quad (3)$$

[a] Department of Mechanical and Aerospace Engineering, Ohio State University, Columbus OH 43210, USA
[b] Department of Materials Science and Engineering, Ohio State University, Columbus OH 43210, USA
[c] Department of Physics, Ohio State University, Columbus OH 43210, USA
* Co-corresponding authors

Here, $v' = {dv}/{dE}$ is the derivative of the sound velocity with respect to the applied electric field and $\gamma$ is the specific heat-weighted mode and wavevector average of $\gamma(i,k)$. From the sound velocity we move to the thermal conductivity using the equation $\kappa = \frac{1}{3}Cvl = \frac{1}{3}Cv^2\tau$, where $C$ represents the volumetric specific heat, $l$ the mean free path, and $\tau$ the relaxation time; $v$ here is the average velocity of the LA and 2 TA modes. At temperatures above the Debye temperature ($\Theta_D$~150-180 K for PMN-PT [4]), the specific heat is constant with $E$, being a function of only the number of atoms per unit volume $n$, $C = 3nk_B$ ($k_B$ is the Boltzmann constant).

In a first approximation, above 200 K, we hypothesize that the scattering time is unaffected by the electric field, based on temperature dependent measurements by Mante and Folger on $BaTiO_3$.[5] There, below 10 K, the phonon mean free path is on the order of the domain size and since this property is field sensitive, the thermal conductivity also depends on electric field. This effect vanishes at higher temperatures where phonon-phonon scattering dominates. Assuming that the relaxation time is independent of electric field ($T \gg 10$ K), we can relate the electric field dependency of both the sound velocity and thermal conductivity $\kappa$ above the Debye temperature as:

$$\frac{\kappa'}{\kappa_o} = 2\frac{v'}{v_0} \quad (4)$$

where again $\kappa' = {d\kappa}/{dE}$. The validity of Equations (3) and (4) was tested experimentally on PZT and confirmed in Wooten et al. [3] without the use of adjustable parameters since $d_{33}$, $d_{31}$, and $\gamma$ were known.

While this approximation holds well for PZT [3], it is more questionable for PMN-PT. Indeed, neutron scattering data on PMN-30PT [2] shows both a strong softening of the lower section of the TA phonon branch and a marked decrease in phonon lifetime. In this case,

$$\frac{\kappa'}{\kappa_o} = 2\frac{v'}{v_0} + \frac{\tau'}{\tau}. \quad (5)$$

The latter term is not attributed to domain wall scattering, but rather to the change in phase space for acoustic phonons to scatter due to changes in the phonon spectrum with electric field. In that sense, the second term in Equation (5) is also due to changes in the phonon spectrum. Experimentally, comparing dynamic measurements of the sound velocity and the thermal conductivity changes with field provides a test for the validity of Equation (4).

If Equations (3) and (4) have predictive value, they allow us to sort amongst known ferroelectrics for the compound that has the largest field dependence of thermal conductivity and is thus most useful in thermal switching applications. The primary application for modulating the heat conduction in FE materials lies in the implementation of voltage-driven heat switches. Heat switches, or thermal transistors, are a particularly useful element in heat management, [6] which is key to the continued scaling up of the complexity in advanced semiconductor integrated circuits. They are also the key element in all-solid-state power generation and refrigeration cycles based on electrocaloric and magnetocaloric effects, where no fluid is circulated but only heat. Many heat switching technologies exist, [7] including one that is based on domain wall motion in ferroelectrics [8] and another on the electrophononic effect based on the lowering of symmetry in ferroelectrics [9–11]; however, the mechanism exploited here, phonon spectrum shifts via electric field, promises to give reasonable effects over a large temperature range if maximized. Ultimately, the goal is to create an electrically modulated, voltage-driven thermal switch with the largest switching ratio.

Equation (3) sheds light on the material parameters needed to maximize the thermal modulation – very anharmonic phonons denoted by a large Grüneisen parameter $\gamma$ and maximization of the summation $(d_{33}+2d_{31})$. The relaxor ferroelectric solid solution system $Pb[Mg_{1/3}Nb_{2/3}]O_3$-$PbTiO_3$ (PMN-PT) is known to exhibit enhanced piezoelectricity. In this work, we investigate the thermal conductivity modulation of PMN-PT by electric field with the specific composition of (*1-x*)$Pb[Mg_{1/3}Nb_{2/3}]O_3$-*x*$PbTiO_3$ with $x$ estimated to be between 0.3 and 0.33 (further abbreviated as PMN-33PT throughout text). The theory anticipated a field dependency an order of magnitude larger and of opposite sign from that in PZT, based on the literature values of $d_{33}$, $d_{31}$ and $\gamma$. In this manuscript, we report this prediction to be experimentally true in a fresh sample, but we surprisingly found that samples that were subjected to multiple cycles of electric field showed an even larger effect and another sign change, ultimately due to changes in the piezoelectric coefficients. Further, we explore the behavior of ${\kappa'}/{\kappa_0}$ over the different crystallographic phases of PMN-33PT with temperature. This paper therefore gives additional proof that ferrons can be propagating acoustic phonons, and thus, by Onsager reciprocity, [12] that a heat-driven polarization flux must exist.

## Methods

### Materials Procurement & Preparation

Single-crystals of PMN-*x*PT ($0.3 \leq x \leq 0.33$) with poling direction of (001) and thickness of 0.5 mm were obtained from MSE Supplies. To apply an electric field, 10 nm of Ti then 90 nm of Au was evaporated onto the primary surfaces in a CHA solution system E-gun evaporator.

### Sound Velocity

To measure the sound velocity, a fresh sample of PMN-33PT with evaporated gold was cleaved into a parallelopiped of approximate size 3 mm × 1 mm. Coiled Cu wires (Omega, diameter of 25 μm) were attached to the electrodes to apply an electric field. DC voltage was applied using a Stanford Research Systems' Data Precision 8200. The sample was placed in a Resonant Ultrasound Spectrometer (RUS) (Alamo Creek Engineering). In the RUS, the sample was subjected to ultrasound frequencies that were amplified if they corresponded to a natural resonant frequency of the entire sample. The peak corresponding to the longitudinal frequency of the compressive wave in the whole sample was selected. This targeted longitudinal resonant frequency was calculated using the detailed sample geometry and dimensions, the Young's modulus, the density provided from manufacturer's website. The shift of the peak frequency, $f$, with applied voltage was tracked by hand.

### Thermal Conductivity

Thermal conductivity measurements were conducted using the static heater-and-sink method on two PMN-33PT samples. Two type T thermocouples consisting of a Cu and Constantan wire were adhered to the first PMN-33PT sample using GE Varnish, which is thermally conducting but electrically insulating. This ensured no short circuiting to the electrodes occurred. On the second sample of PMN-33PT, StyCast epoxy was used instead of GE varnish which is also thermally conducting but electrically insulating. Prior experiments established that there is little difference between measurements made on samples mounted with either StyCast or GE varnish, but StyCast is mechanically stronger. Sample 1 was subsequently positioned on an alumina base, serving as a heat sink, and secured with GE varnish, while sample 2 was affixed using StyCast. A strain gauge was placed with GE varnish on sample 1 and with StyCast on sample 2 acting as the heater. The samples were then placed individually in a vacuumed environment inside a Lakeshore nitrogen-cooled cryostat. Sample 1 (labeled "fresh") was measured directly after mounting, while sample 2 (labeled "cycled") was cycled multiple times through its hysteresis loop at the various temperatures. The temperature was allowed to stabilize for at least one hour prior to each measurement. During both measurements, the electric field was maintained for about 5 minutes before the thermal conductivity was recorded to alleviate any pyroelectric artifacts.

The absolute value error of thermal conductivity near room temperature in the static heater and sink method is on the order of 10%. It is dominated by the radiative heat losses and by geometrical uncertainties. The latter is typically dominated by the distance between thermocouples, where the error is given by the ratio between the size of the thermocouple contacts (which gives the uncertainty) and the distance between them. Here, this uncertainty is calculated to be 1.4-1.8%. In the cryostat, the radiative heat losses correspond to a thermal conductance that has been measured using a standard iron bar of known thermal conductivity; they are 0.10 mW $K^{-1}$ at 200 K, 0.44 mW $K^{-1}$ at 300 K and 0.85 mW $K^{-1}$ at 360 K. Given the thermal conductivity and the geometry of the PMN-33PT samples, the losses represent <7% at 200 K, <29% at 300 K but <55% at 400K. These losses have been subtracted from the measured conductance, but that procedure carries an uncertainty due to the difference in emissivity between the sample and the iron standard, which itself has an error of the order of 30%. Taken together, the radiative losses represent an uncertainty in absolute value of 2% at 200K, 10% at 300K, but up to 18% at 360 K. The plots shown in Figures 4, S2 and S3 represent the field-dependence of the conductivity, which is only affected by the relative error. This is much smaller and dominated by the noise on the thermocouples, equivalent to about 10 mK or 1% of the temperature difference and thus the measured conductivity.

### Transverse Piezoelectric Coefficient $d_{31}$

To determine the $d_{31}$ coefficient, we tracked the displacement of the top surface of the sample as a function of applied AC field. The sample measuring approximately 10 mm × 10 mm × 0.5 mm was securely affixed to a microscope glass slab using GE varnish. As the top surface was smaller than the laser beam size, a piece of Kapton tape was placed atop the sample. This layer provided a broader area for the laser-beam to hit the edge of the sample ensuring smaller measurement error by increasing the reflected signal. For this displacement measurement, we relied on the Keyence LKG32 Laser Sensor. To apply the AC voltage, we employed the Techron 5050 Linear Amplifier in tandem with the Agilent 33120A Waveform Generator. Throughout the experiment, we continuously monitored the changes in displacement using the Rigol MSO5074 oscilloscope.

A parallel experiment was carried out using the cycled sample. The objective of this second experiment was to elucidate the impact of cycling the sample through several sweeps of the electric field. The mounting of the sample for this experiment was slightly different from the first measurement. The laser beam hit the strain gauge placed on the top of the sample to measure the longitudinal displacement while applying electric field. The $d_{31}$ coefficient was calculated by taking the slope of strain vs electric field.

### Longitudinal Piezoelectric Coefficient $d_{33}$

To measure the $d_{33}$ coefficients, we used an MTI 2100 Fotonic measurement system while applying a DC voltage via a Stanford Research Systems' Data Precision 8200. The MTI 2100 Fotonic system utilizes fiber optics to precisely measure the small displacement parallel to the applied electric field. A fresh sample equipped with evaporated gold electrodes was placed inside a Poly K test fixture consisting of a metallic sample holder acting as electrical conductors on the sample electrodes and a shiny metallic surface where the fiber optic light was reflected and output as a measured displacement. Both sample holder and metallic surface are attached to micrometers allowing for precise positioning and instrument calibration. The fresh sample $d_{33}$ coefficient was calculated by taking the slope of strain vs electric field and restricting the field to stay within the linear regime ($|E|\leq 2\times10^{5}$ V m$^{-1}$). The sample was then subjected to roughly 20 cycles of electric field and the displacement tracked as a function of voltage. The $d_{33}$ coefficient was calculated by taking the slope of the strain as a function of field. These measurements were repeated on three samples to test for reproducibility.

## Results

The compound we used here was PMN-33PT ($0.3\leq x \leq 0.33$) in single-crystal form. X-ray diffraction analysis showed a lattice constant of 4.026 Å at room temperature (see Figure S1 in supplementary information and materials). The material goes through several crystallographic phase transitions as a function of temperature and electric field. [13] At zero field, it is tetragonal (T) above about 320 K and monoclinic (M) below. The M phase contains two sub-phases, $M_A$ and $M_C$, where the subscript denotes different polarization vector directions. The phases are readily apparent in thermal conductivity measurements at zero field with the heat flux applied along the (100) direction, reported in Figure 1. Below ~225 K, PMN-33PT exists in the $M_A$ phase, transitioning to the $M_C$ phase between ~225 K and ~325 K. Beyond ~325 K, it adopts a T structure. Notably, as no electric field is applied at this time, the observed phase transitions solely occur because of temperature variations.

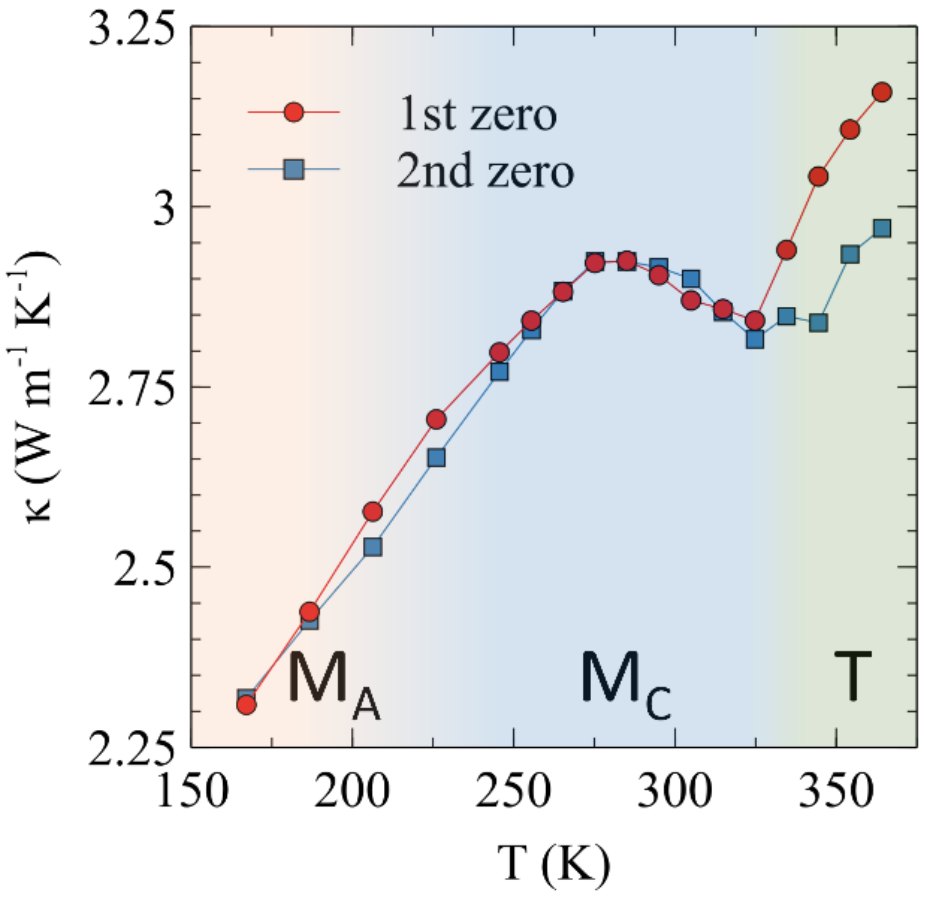

Figure 1: Temperature dependent phase transitions observed in thermal conductivity measurements at zero field taken during field sweeps. In the legend, ‘1st zero’ refers to the thermal conductivity value at zero field as the swept field is decreasing, and ‘2nd zero’ refers to the value at zero field as the swept field is increasing (see later, Fig. 4). $M_A$, $M_C$, and T refer to the two monoclinic and tetragonal phases that PMN-33PT passes through as the temperature changes.

These phases are also dependent on the electric field. For example, if the sample is slightly above room temperature, it can transition from $M_A$ to $M_C$ with increasing field. These transitions are obvious in the strain vs electric field curves found in Davis et al. [13] Ultimately, these transitions are a function of temperature, electric field, synthesis protocols, composition, and poling conditions. [13,14] For example, PMN-33PT has a large piezoelectric coefficient ~$d_{33}$=1500 pC N$^{-1}$; [15] a recent paper [14] related the poling conditions to $d_{33}$ and found that $d_{33}$ can change drastically with the poling electric field strength and AC or DC poling conditions. Tachibana et al. [4] reported a lower thermal conductivity for PMN-32PT, suggesting that thermal conductivity is highly sensitive to sample conditions and thermal stresses induced by cycling, including the growth method; a higher thermal conductivity is indicative of a more pristine sample. These materials are complex in structure which leads to a considerable spread in reported absolute values of $\kappa$.

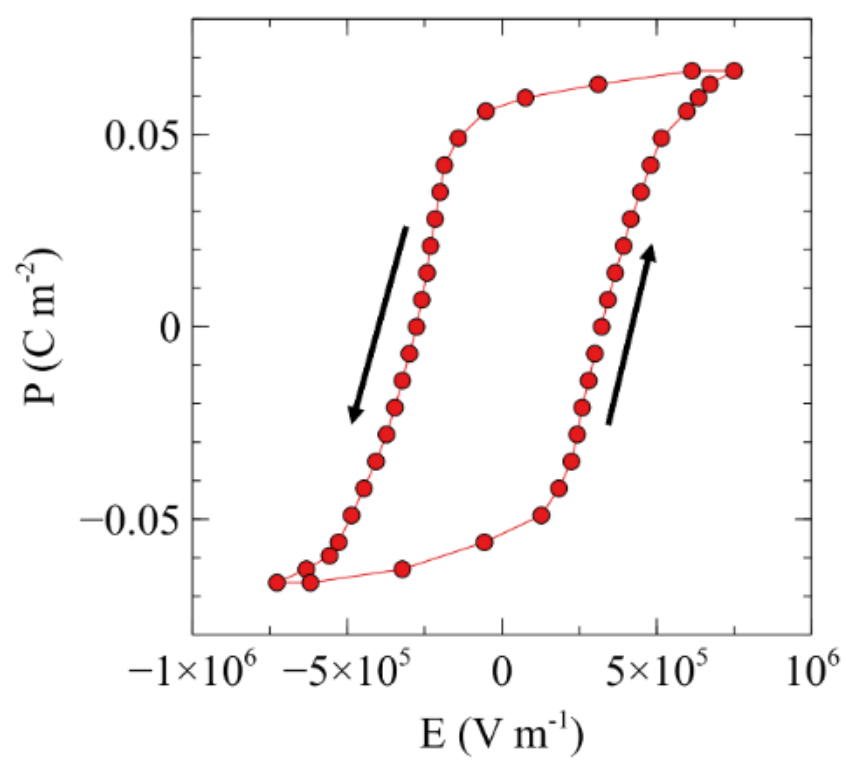

Figure 2: Polarization of PMN-33PT at room temperature. Arrows indicate the direction of the swept field.

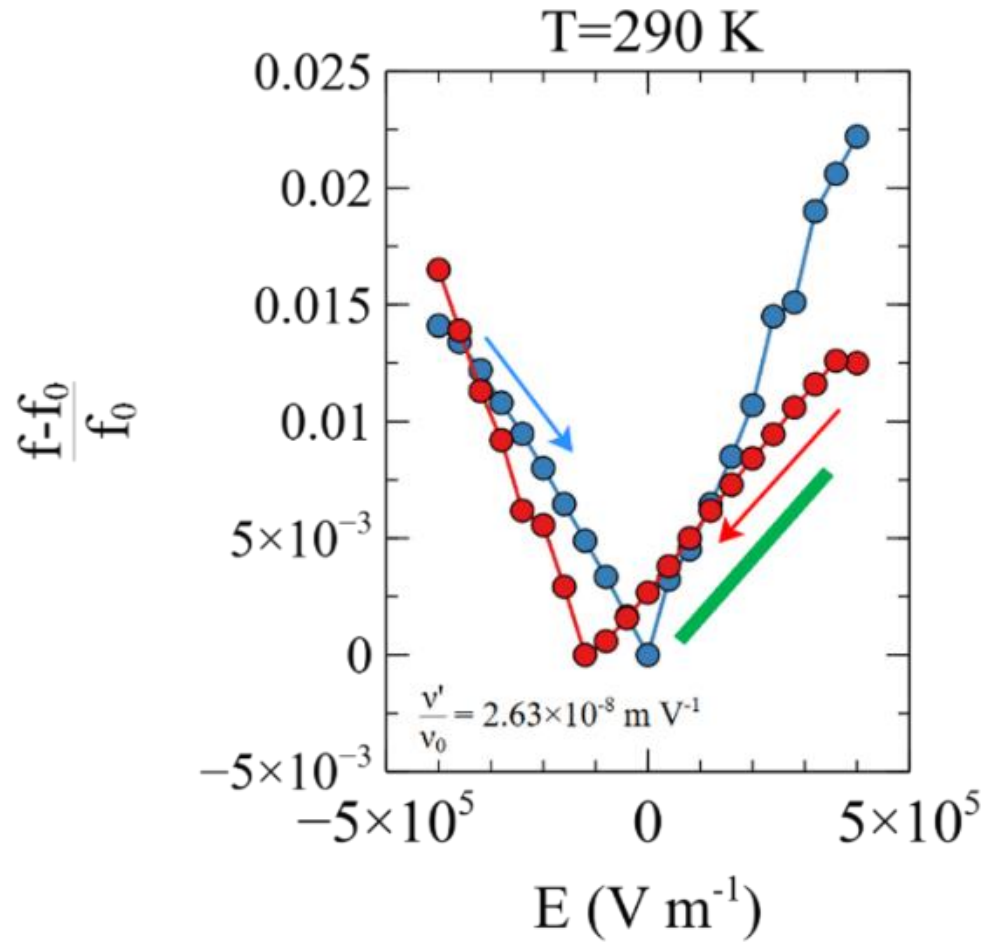

Figure 3: Resonant Ultrasound Spectroscopy (RUS) measurement of the frequency of the resonance of the longitudinal compressive mode of a parallelepiped sample of PMN-33PT at 290 K. The relative change of this longitudinal resonant frequency is tracked as a function of electric field. This relative change of frequency equals the relative change of $v'/v_0$ of the longitudinal acoustic phonon along the (100) direction of the crystal. The green line indicates where the slope was taken to obtain the value of $v'/v_0$.

The polarization vs electric field along the (001) direction results at room temperature are shown in Figure 2. The coercive field is about 3×10$^5$ V m$^{-1}$ and the saturation field is about 6×10$^5$ V m$^{-1}$ with a polarization of about 0.06 C m$^{-2}$.

Figure 3 shows the relative change with electric field of the resonant frequency, $f$, that corresponds to the longitudinal compressive wave of the whole sample in the RUS. Because both $f$ and the sound velocity of the LA phonon depend on the square root of the elastic constant $c_{11}$, this relative change equals the relative change in the sound velocity of the longitudinal acoustic phonon, i.e. $\frac{f - f_0}{f_0} = \frac{v'_{LA}}{v_{LA,0}}$. These measurements could not be taken on the cycled sample, as it had been instrumented for thermal conductivity measurements and its instrumentation perturbed the natural resonant frequencies of the sample.

Figure 4a and b illustrate the behavior of the thermal conductivity of the fresh and cycled PMN-33PT samples with field. The first run gives a $\kappa'/\kappa_0$ of almost 4×10$^{-8}$ m V$^{-1}$. The thermal conductivity's field dependency changes during subsequent cycles of positive and negative field applied to the sample as shown in Figure S2 which illustrates the first four cycles. The second sample cycled 20 times near room temperature (Figure 4b) gives $\kappa'/\kappa_0$ = -3.86×10$^{-7}$ m V$^{-1}$. This condition is stable and reproducible (the curve after 10 cycles is shown in the inset and differs very little from the one cycled 20 times). The electric field dependence of the thermal conductivity shows a sign flip and is an order of magnitude larger vis-à-vis the fresh sample. The value is now two orders of magnitude larger than that reported in PZT ($\kappa'/\kappa_0$= -7×10$^{-9}$ m V$^{-1}$). [3] Considering the minimum and maximum thermal conductivity values, the change in thermal conductivity of the cycled PMN-33PT sample with field at room temperature is close to 8%. This increases to 11% at 364 K (see Figure S3).

Figure 5 illustrates $d_{31}$ measurements on two distinct samples: one pristine, untouched by an electric field, and the other, the cycled sample previously employed for thermal conductivity measurements. To measure the transverse piezoelectric constant, we measured the change of the length of the sample along the (100) direction as a function of the electric field along the (001) direction. This change in the length of the sample corresponds to changes in the strain tensor; the derivative of these strain tensor components gives the piezoelectric coefficient. We calculated a $d_{31}$ of -1327 pC N$^{-1}$ for the fresh sample and -476 pC N$^{-1}$ for the cycled sample, a reduction of nearly three times.

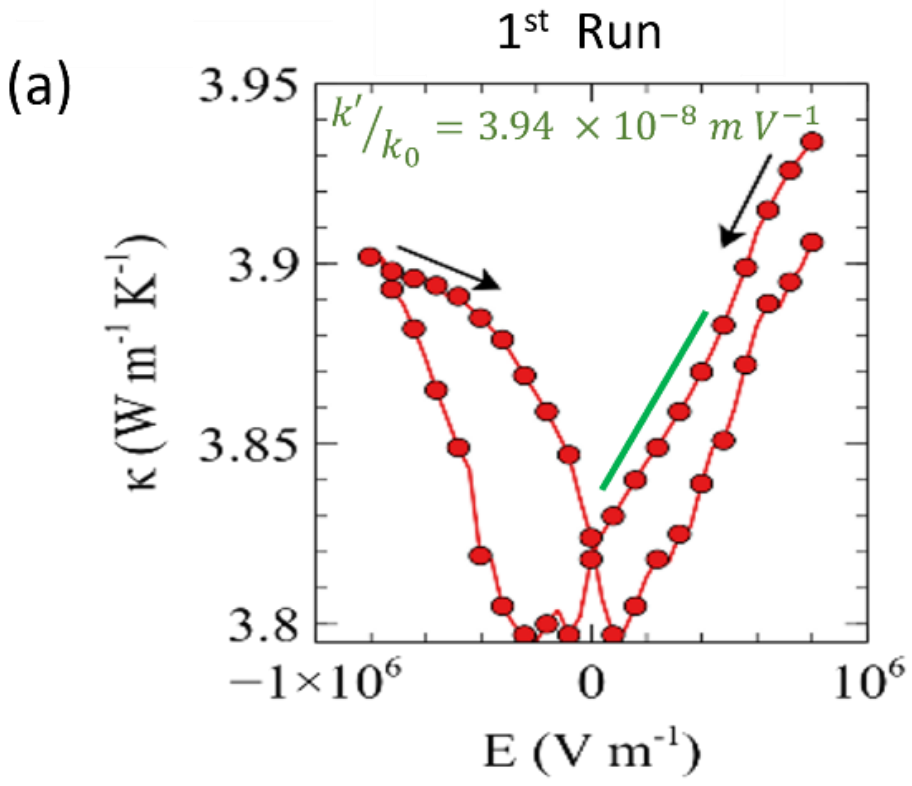

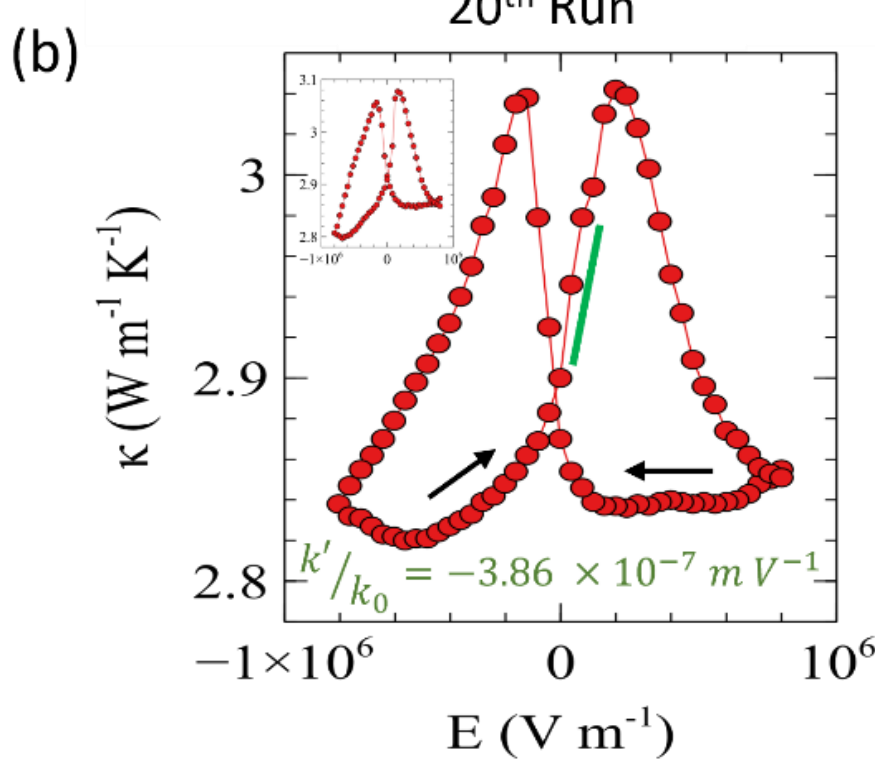

Figure 4: a-b) Room temperature thermal conductivity measurements on (a) a fresh sample on its first run and (b) a cycled sample on its 20th run. The (b) inset gives the data at the 10th run. This is essentially the same as at the 20th run, showing that while a transition occurs during the first 10 runs (see Figure S2), the sample is stable after about 10 cycles with electric field. The arrows indicate direction of sweep while the green line represents where the slope was taken to calculate the $\kappa'/\kappa_0$ values displayed below each plot.

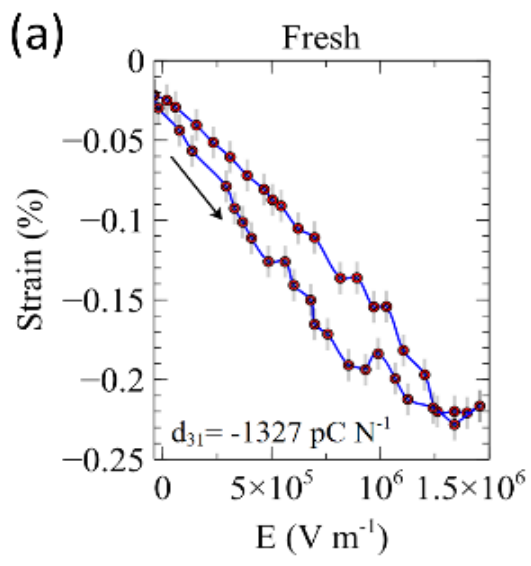

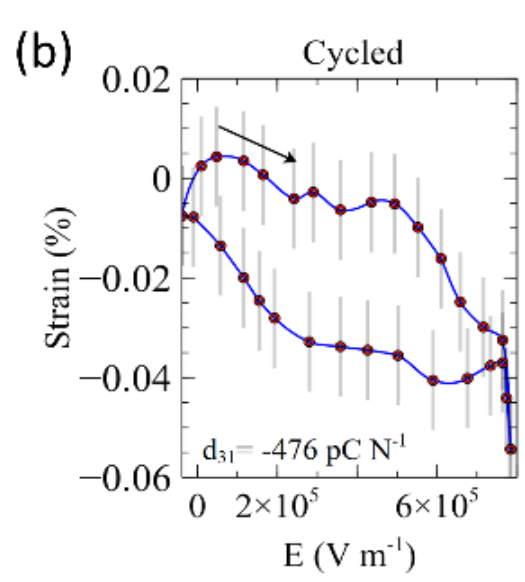

Figure 5: a-b) Transverse piezoelectric strain measurements on (a) a fresh PMN-33PT sample and (b) a cycled PMN-33PT sample. The slopes taken in decreasing field, indicated with arrows, were used to calculate the piezoelectric constant $d_{31}$ values.

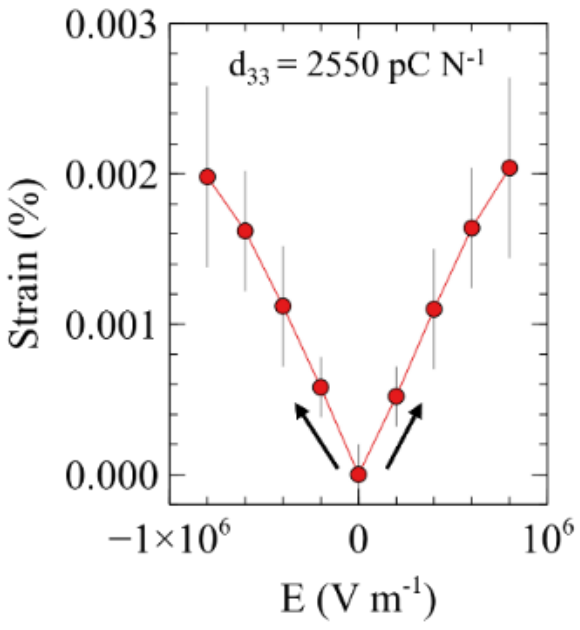

Figure 6: Longitudinal piezoelectric measurement on a sample of PMN-33PT cycled ~20 times. The slope was averaged between the negative and positive fields to obtain a $d_{33}$ value of 2550 pC N$^{-1}$ for the cycled sample.

We also investigated the longitudinal piezoelectric coefficient - $d_{33}$, corresponding to the displacement along the (001) direction parallel to the applied field. We measured $d_{33}$ on a fresh sample by alternating between 0 and +100 V, back to zero to ensure we remained in the linear regime and then to -100 V. This gave us a $d_{33}$ of 1550 pC $N^{-1}$ for the fresh PMN-33PT. We then cycled this sample roughly 20 times electrically. Figure 6 shows the results for the cycled sample in which we obtained $d_{33}$ = 2550 pC $N^{-1}$ by averaging the slope in the positive and negative fields. Note that the zero-field value was zeroed out to account for hysteretic effects.

## Discussion

Table I summarizes the values for $d_{33}$, $d_{31}$, $v'/v_0$ and $\kappa'/\kappa_0$, to be tested against Equations (3) and (4), with a Grüneisen parameter $\gamma$=15 for PZT; [16] the same value will be assumed here for PMN-PT since $PbTiO_3$ is presumably the active ingredient. The measured $d_{31}$ from Figure 5a and $d_{33}$ measured in the linear regime can be used to calculate the predicted values of $v'/v_0$ and $\kappa'/\kappa_0$ which correspond to the fresh sample depicted in Figures 3 and 4a. The results, in the column labeled "Predicted" in Table I, are quite close to the experimental values without the use of any adjustable parameter. Furthermore, Equation (4) holds approximately for the fresh sample, corroborating the hypothesis that $\tau$ has little field dependence. The initial results on fresh PMN-33PT are quantitatively explained by Equations (3) and (4) without adjustable parameters, just as the case for PZT. [3] Comparing the sound velocity and thermal conductivity results from Figures 3 and 4a to that of PZT, we do indeed see the resulting curves of PMN-33PT invert, and that the latter are an order of magnitude larger. The summation of the piezoelectric coefficients $(d_{33} + 2d_{31})$ of PMN-33PT has the opposite sign than that of PZT, [3] leading to the sign inversion in the sensitivity of sound velocity and thermal conductivity to field. This adds definitive support to the theory [3,12] that the dispersion of acoustic phonons is affected by an external electric field and further, that acoustic phonons can thus be considered as ferrons carrying polarization, since at finite values for the phonon momentum they are coupled by piezoelectric strain to the optical phonons.

This treatment can be extended to the stable trace reported on the cycled sample in Figure 4b, using the $d_{31}$ value of -476 pC $N^{-1}$ calculated from the measurement shown in Figure 5b and $d_{33}$ value of 2550 pC $N^{-1}$ calculated from the measurement shown in Figure 6. The sign inversion between cycled (Figure 4b) and non-cycled (Figure 4a) PMN-33PT is also predicted by the change in piezoelectric coefficients but not the additional increase in sensitivity by an order of magnitude.

We attribute this order of magnitude discrepancy between predicted and measured $\kappa'/\kappa_0$ to the dependence of $\tau$ on the electric field (Equation 5) observed in PMN-30PT. [2] A detailed study of the effect of scattering on the thermal conductivity using phonon lifetimes from inelastic neutron scattering is left for future work, as it has the potential to further enhance $\kappa'/\kappa_0$ and reach more practical thermal switching ratios.

To be complete we must consider competing theories. Negi et al. [14] argue that a change in thermal conductivity in PMN-PT can come from domain wall scattering. In principle, there must be a contribution of this mechanism to the electric field dependence of $\kappa$, but we argue that the principal mechanism arises through a change in sound velocity and phonon spectrum. The first reason for this argument is the RUS measurements shown in Figure 3, where we show unequivocally that the sound velocity depends on the electric field; sound velocity is a property at thermodynamic equilibrium and not a transport property. The second reason is that the domain sizes in Negi's work are of the order of 1 µm, much larger than the average phonon mean free path. While domain wall scattering has been shown unequivocally to cause an electric field dependence on the thermal conductivity at cryogenic temperatures [5], this effect vanishes as the temperature increases to values where the phonon thermal mean free path becomes much smaller than the domain wall size, typically above 20 K.

## Conclusions

In summary, our study involved measuring thermal conductivity and sound velocity of PMN-33PT on two samples, one fresh and one cycled. The fresh sample exhibited values consistent with theoretical predictions based on the ferron theory and specific piezoelectric coefficients. The cycled sample, subjected to a sweeping electric field 20 times, displayed a sign change in $\kappa'/\kappa_0$ compared to the results obtained in the fresh sample and an increase by an order of magnitude. This suggested alterations in piezoelectric coefficients, as well as a possible contribution of an electric field dependent relaxation time, probably induced by a change in the optical phonon modes. This would result in a much larger phase space for scattering of the acoustic modes. Interestingly, the summation $(d_{33} + 2d_{31})$ exhibited differing signs and values between PZT and PMN-33PT, and then a second sign change between the electrically cycled and the fresh sample of PMN-33PT. Calculated $\kappa'/\kappa_0$ values using measured $d_{31}$ and $d_{33}$ aligned with expected piezoelectric coefficient changes outlined in literature. [14] Though a magnitude difference in the cycled sample may relate to using samples with different histories and suspected electric field dependent scattering, further investigation is necessary. Comparing PMN-33PT with Wooten et al.'s [3] findings for PZT indicates an increase in field sensitivity of the thermal conductivity by one order of magnitude for the fresh PMN-33PT sample and two orders of magnitude for the cycled sample, suggesting promising applications despite the need for further signal enhancement in studies on thermal switches.

Table I: Material type, properties, predicted and measured $v'/v_0$ and $\kappa'/\kappa_0$ values for PZT, fresh PMN-33PT, and cycled PMN-33PT. All predicted values were calculated with $\gamma$ = 15 [16].

| | | | | PREDICTED | MEASURED | |
|---|---|---|---|---|---|---|
| MATERIAL | TYPE | $d_{33}$ [pc $N^{-1}$] | $d_{31}$ [pc $N^{-1}$] | $-\gamma(d_{33} + 2d_{31})$ [m $V^{-1}$] | $v'/v_0$ [m $V^{-1}$] | $\kappa'/\kappa_0$ [m $V^{-1}$] |
| PZT | FE | 435 [17] | -100 [18] | $-3.5\times10^{-9}$ [3] | $-3.3\times10^{-9}$ [3] | $-3.3\times10^{-9}$ [3] |
| PMN-33PT (fresh) | RFE | 1350 | -1327 | $2.0\times10^{-8}$ | $2.6\times10^{-8}$ | $3.9\times10^{-8}$ |
| PMN-33PT (cycled) | RFE | 2550 | -476 | $-2.4\times10^{-8}$ | - | $-3.9\times10^{-7}$ |

## Author contributions

DR was responsible for data curation, formal analysis, investigation, and writing – original draft and review and editing. BLW was responsible for data curation, formal analysis, investigation, visualization, supervision, and writing – original draft and review and editing. JPH was responsible for conceptualization, funding acquisition, supervision, and writing – original draft and review and editing.

## Conflicts of interest

There are no conflicts to declare.

## Data availability

Data available upon request.

## Acknowledgements

The primary funding for this work (DR, JPH) is from the US National Science Foundation grant CBET 213 3718. BLW is supported by a US Department of Defense SMART fellowship. The authors thank the OSU Smart Materials and Structures Lab, specifically Marcelo Dapino who assisted in $d_{31}$ measurements. The authors thank Prof. Patrick Woodward for engaging discussion involving relaxor ferroelectric materials. Metallization was conducted at OSU's Nanotech West Lab; the authors thank the lab manager, Paul Steffan, for the detailed training and guidance. Lastly, the authors thank Mohamed Nawwar for his assistance with XRD analysis.

# Supplementary Information & Materials: Electric field dependent thermal conductivity of relaxor ferroelectric PMN-33PT through changes in the phonon spectrum

Delaram Rashadfar,[a] Brandi L. Wooten [b,*] and Joseph P. Heremans [a,b,c,*]

*[a.] Department of Mechanical and Aerospace Engineering, Ohio State University, Columbus OH 43210, USA*
*[b.] Department of Materials Science and Engineering, Ohio State University, Columbus OH 43210, USA*
*[c.] Department of Physics, Ohio State University, Columbus OH 43210, USA*
** Co-corresponding authors*

Figures S1a-e show the x-ray diffraction pattern for an unpoled and a poled sample of PMN-33PT. The sample was x-rayed using a Bruker D2 Phaser and then coated with silver paint to apply electrodes. It was subjected to a field of $8\times10^5$ V $m^{-1}$ for about an hour. The paint was removed with acetone and the sample was x-rayed again. Using Rietveld refinement and a silicon standard for sample displacement, a lattice constant of 0.402607 nm was calculated for the unpoled sample and 0.402649 nm for the poled sample giving a change of $4.2\times10^{-5}$ nm and strain of -0.01%. The smaller unidentified peak around ~39.7° in Figure S1c could be a (111) peak and hints at the inclusion of a secondary phase; however, the (222) peak estimated to be around 83.2° is not present, so this remains inconclusive.

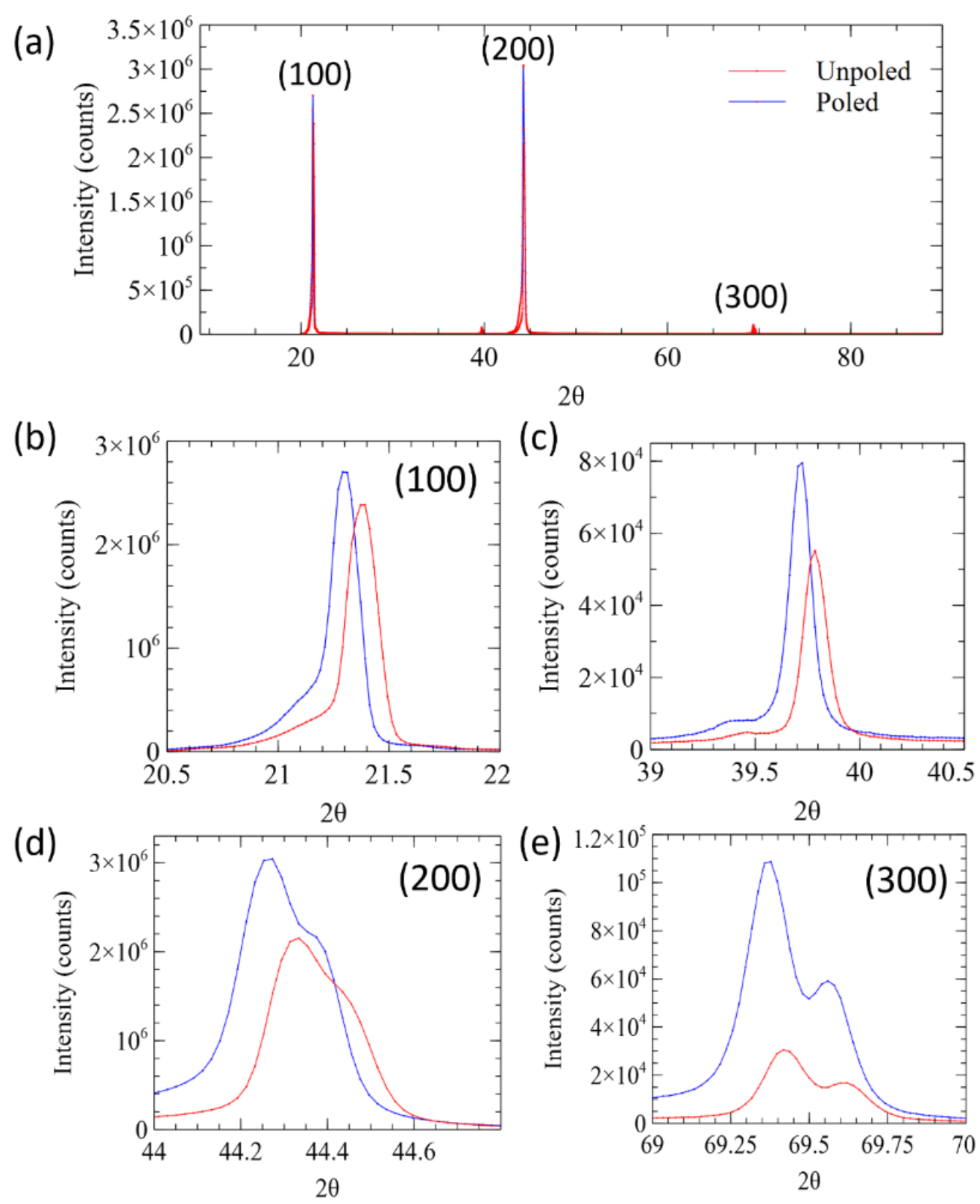

Figure S1. a-e) X-ray diffraction patterns for a fresh unpoled sample of PMN-33PT and a poled sample that experienced ~1 hour under an electric field of $8\times10^5$ $Vm^{-1}$. Zoomed in plots for the peaks labeled (100), (200), and (300) are provided in (b), (d), and (e) respectively. The smaller peak at ~39.7° in (c) could be a (111) peak; however, the (222) peak estimated to be around 83.2° is not present.

**Figures S2a-d** display the thermal conductivity of the first PMN-33PT sample near room temperature on the first, second, third, and fourth sweep of the electric field. The experimental values of the thermal conductivity dependency are below each plot. The point worth noting is that they are all in the same order of magnitude, but $\kappa'/\kappa_0$ begins to decrease by the fourth sweep in the electric field signaling a transition. The formation of a peak is also visible by the 4th run.

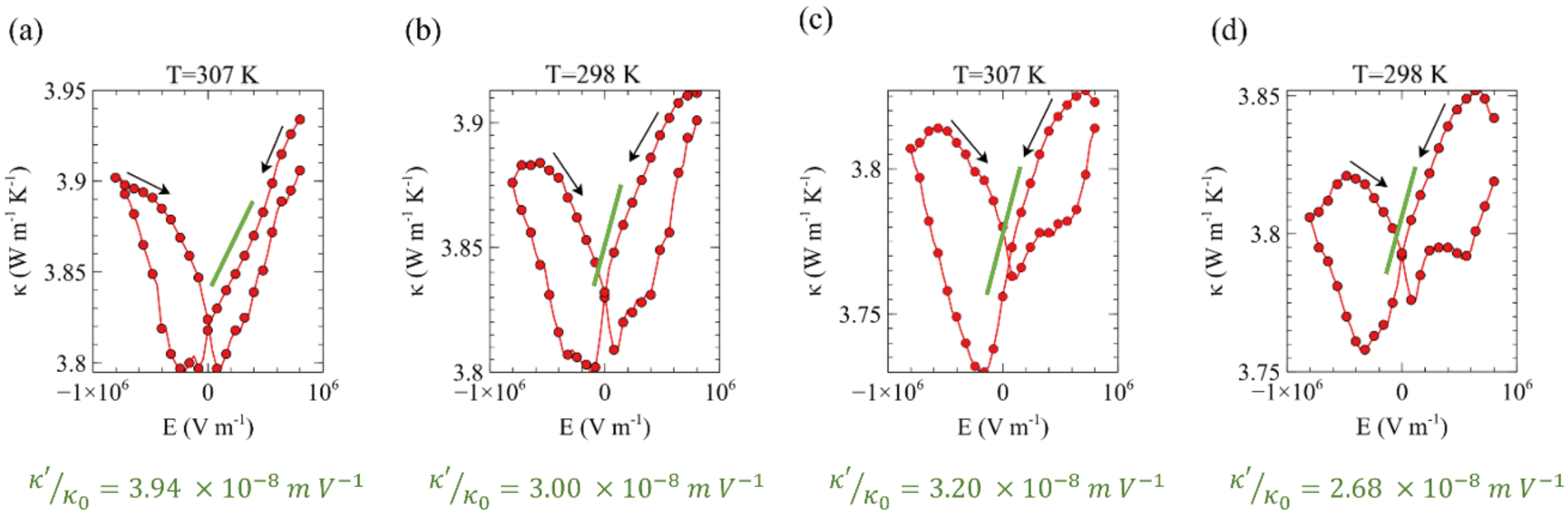

**Figure S2.** a-d) Thermal conductivity measurements around room temperature on a fresh PMN-33PT sample: (a) the first run, (b) the second run, (c) the third run, and (d) the fourth run of sweeping field. The arrows indicate the direction of sweep and the green line indicates where the slope was taken to calculate the values shown in **Table S1**.

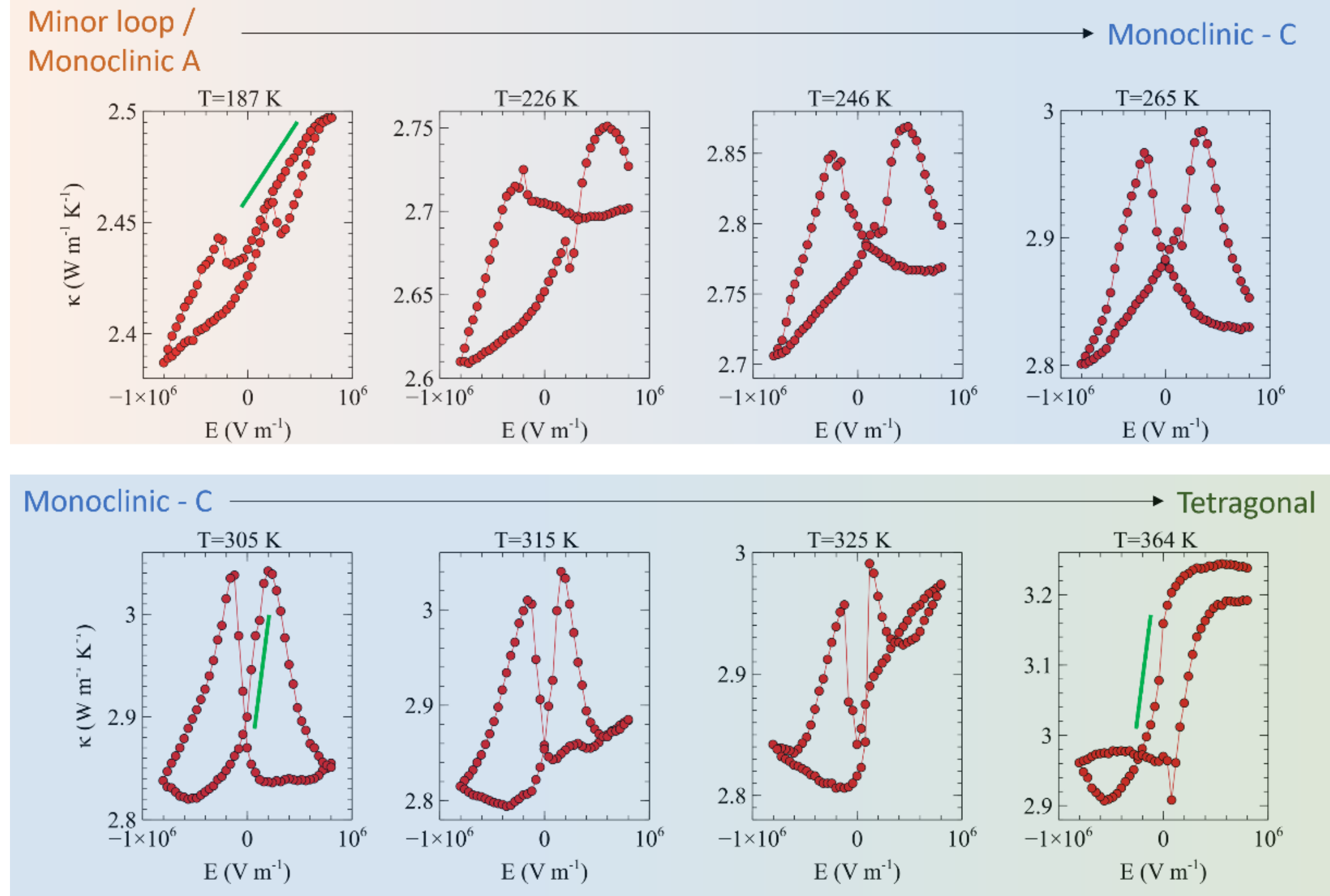

**Figure S3.** Thermal conductivity measurements at different temperatures on the cycled PMN-33PT sample under electric field. At cooler temperatures, we are in the monoclinic A phase; however, we may also be on a minor loop as the field may not be strong enough at this temperature to reach saturation. Around 265 to 315 K, we are in the monoclinic C phase, and we see symmetric curves with loop closure. Upon heating up to 364 K, we move to the tetragonal phase. Here, we see the greatest switching ratio. The green lines in the *T*=187, 305, and 365 K plots indicate where the slope was taken to calculate the values shown in **Table S1**.

**Figure S3** shows how thermal conductivity changes as a function of electric field at different temperatures on the cycled sample. At the lower temperature *T*=187 K, we see an odd function of field but with hints of peaks at the coercive field. This observed trend could be indicative of a distinct phase; however, it is likely that we are on a minor loop of the hysteresis curves and cannot safely apply enough voltage to saturate the polarization in the experiment. As the temperature increases, broader peaks begin to form at the coercive field values, until we see an even function of field at 265 K. This trend persists until about 315 K. Upon further heating, the thermal conductivity shifts to an odd function of field with observed saturation at the higher fields.

**Table S1.** Collected $\kappa'/\kappa_0$ and conductivity ratios $\kappa_{max}/\kappa_0$ values from a fresh sample with the first four runs around room temperature and a cycled sample at various temperatures.

| Sample | Phase | Temperature [K] | $\kappa'/\kappa_0$ [m V$^{-1}$] | $\kappa_{max}/\kappa_0$ |
|---|---|---|---|---|
| Fresh – Run 1 | $M_C$ | 307 | $3.94 \times 10^{-8}$ | 1.030 |
| “Fresh” – Run 2 | $M_C$ | 298 | $3.00 \times 10^{-8}$ | 1.019 |
| “Fresh” – Run 3 | $M_C$ | 307 | $3.20 \times 10^{-8}$ | 1.022 |
| “Fresh” – Run 4 | $M_C$ | 298 | $2.68 \times 10^{-8}$ | 1.016 |
| Cycled | $M_A$ (Minor loop) | 187 | $-3.13 \times 10^{-8}$ | 1.024 |
| Cycled | $M_C$ | 305 | $-3.86 \times 10^{-7}$ | 1.060 |
| Cycled | T | 364 | $-2.61 \times 10^{-7}$ | 1.026 |